\documentclass{arxivIEEEcsmag}

\usepackage[colorlinks,urlcolor=blue,linkcolor=blue,citecolor=blue]{hyperref}

\usepackage{upmath}

\jvol{XX}
\jnum{XX}
\paper{8}
\jmonth{April/May}
\jname{CiSE}
\pubyear{2023}

\begin{document}


\title{Research Software Engineers: Career Entry Points and Training Gaps}

\author{Ian A. Cosden}
\affil{Research Computing, Princeton University, Princeton, NJ, 08544 USA}

\author{Kenton McHenry}
\affil{NCSA, University of Illinois at Urbana-Champaign, Urbana, IL 61801 USA}

\author{Daniel S. Katz}
\affil{NCSA \& CS \& ECE \& iSchool, University of Illinois at Urbana-Champaign, Urbana, IL 61801 USA}

\markboth{CiSE: Special Issue on the Future of Research Software Engineers in the US}{Paper title}

\begin{abstract}
As software has become more essential to research across disciplines, and as the recognition of this fact has grown, the importance of professionalizing the development and maintenance of this software has also increased. The community of software professionals who work on this software have come together under the title Research Software Engineer (RSE) over the last decade. This has led to the formalization of RSE roles and organized RSE groups in universities, national labs, and industry. This, in turn, has created the need to understand how RSEs come into this profession and into these groups, how to further promote this career path to potential members, as well as the need to understand what training gaps need to be filled for RSEs coming from different entry points. We have categorized three main classifications of entry paths into the RSE profession and identified key elements, both advantages and disadvantages, that should be acknowledged and addressed by the broader research community in order to attract and retain a talented and diverse pool of future RSEs.

\end{abstract}

\maketitle

\chapterinitial{Software} has grown as a key part of research along with digital computers, since the 1940s. 
Software developers originally came from the mathematics field, learning to program as needed, and then as computer science and software engineering developed, these fields began to develop standard curricula and training, leading to professional software development practices. 
While the initial software developers came from a research environment, in part because all computing was initially research, as computing became more common and more widely used in business, programming and software engineering also became more formalized and professionalized. 
This led to a dichotomy between software engineering as taught and used in business settings, where it has been a profession, and as used in research, where it has often been one task among many performed by researchers, typically in universities and national laboratories, but also in industry in some cases.

In 2012, the lack of a professional role for software developers in research came to a head, and a breakout session in the 2012 Collaborations Workshop \cite{baxter} focused on common issues found by such software developers. 
The work done in this session and shortly thereafter created and defined the terms Research Software Engineer and Research Software Engineering, and began a movement that has now grown into a community with almost 10000 members globally, multiple annual conferences and workshops, at least 9 national and multinational associations and societies, and formal RSE groups and career paths in many universities and national laboratories.

The authors of this article represent co-founders of the US Research Software Engineer Association (US-RSE) \footnote{ \href{https://us-rse.org}{https://us-rse.org}} and two of the largest RSE groups in the US, 32 RSEs at NCSA at the University of Illinois at Urbana-Champaign \cite{katz-1, katz-2}, and 18 RSEs at Princeton \footnote{ \href{https://researchcomputing.princeton.edu/services/research-software-engineering/group-members}{https://researchcomputing.princeton.edu/services/research-software-engineering/group-members}}. These three activities have seen tremendous growth: US-RSE started in 2018 and now has almost 1300 members, the RSE team at NCSA Software Development team had 5 members in 2012 and now has 32, and the central RSE group at Princeton started in late 2016 with a single RSE and will grow to 28 full-time staff 2023 \cite{princeton-rse-expansion}. As leaders of these efforts, we also see a number of challenges, including a lack of general awareness of:

\begin{itemize}
    \item RSEs and RSE issues from stakeholders outside the RSE community, including many university administrators, many research funders, and many research publishers. 
    \item Possible RSE roles by potential RSEs, including secondary students, undergraduates in both computer science and other fields, and graduate students in computer science and in computation and data-focused disciplines
\end{itemize}

While there has been progress in raising awareness in these groups over the past ten years, the authors estimate based on their experience that the awareness is at best about 5\% for some groups such as funders, and probably much smaller for potential RSEs.

This leads to two issues that diminish the amount and quality of software used in research:

\begin{itemize}
    \item The lack of general awareness and support disincentivizes those who are RSEs and are trying to grow RSE groups.
    \item The lack of awareness by potential RSEs leads to lower than desired pools of candidates for existing positions, and when potential RSEs do discover this career path, they often aren't well prepared for these positions.
\end{itemize}

Because, as of now, there are no formal educational programs specifically designed to prepare RSEs to enter the profession, RSEs emerge from multiple demographics. 
In the next section of this article, we distinguish between a number of different entry points for different groups of potential RSEs, and discuss associated challenges, including awareness and preparation.

\section{RSE Career Entry Points}
The term “Research Software Engineer” can be used to broadly classify individuals who use software engineering to advance research. 
This broad classification means that a qualified candidate might have many different educational backgrounds and experiences. 
It also means, however, that the specifics of a particular position often substantially define the requirements of a role. 
Nevertheless, a handful of overarching categories tend to emerge, each with their own strengths and weaknesses.
We want to highlight that because the RSE profession is still relatively new there are a number of challenges associated with each entry point, but these shouldn't be seen as outweighing the advantages that we have clearly witnessed.
Rather, our hope is to identify the key challenges that current entrants to the profession are likely to face in an effort to focus community efforts on minimizing them.
By providing avenues to reduce these challenges we stand to improve and diversify the RSE pipeline and better retain existing RSEs.

We will discuss the three top level categories that capture most RSE experiences: a domain science background, a pure computer science background, and industry experience as a software developer. 
We also recognize that there will be cases where an individual might fall into two categories simultaneously, for example, a physics PhD graduate who worked in industry for three years as a software developer before taking their first RSE position. 
These special cases can ameliorate some of the challenges associated with one of the categories individually, but won't be specifically addressed here.

\subsection{Domain Science}
According to the 2022 RSE survey \cite{hettrick}, over 75\% of US respondents identified with an educational background of something other than computer science. 
Similarly, over 50\% of all respondents had listed a PhD as their highest level of education. Clearly formally trained researchers in domain sciences is one of the largest entry points to the RSE profession. 
This demographic is consistent with the authors' observations. 
In this category we include both recent graduates and those with additional experience, either in a non-academic setting or as a researcher, such as a postdoc. 
Often, these RSEs are people who started in a particular discipline but discovered they enjoyed the software aspects more than other aspects, and possibly wanted to broaden the application of their software skills.

Entering the RSE profession with a domain science and recent research background (i.e., as a recent PhD graduate or postdoctoral researcher) has a number of benefits that are immediately applicable to an RSE role. 
First, the new RSE has sufficient understanding to digest new research problems and communicate effectively with non-RSE researchers. 
Second, they understand the research culture, goals, and incentives, making new collaborations with researchers at times easier as they speak a shared language. 
Additionally, those self-selecting to enter the RSE professional have typically learned software development skills independently, thus developing the capacity to self-learn new technologies and approaches.

RSEs with a research domain background potentially interested in an RSE or RSE-like role will often face four main challenges: (1) awareness, (2) technical preparation, (3) career prospects, and (4) crossing disciplines.

Because the RSE name and career path has only recently seen an increase in exposure and publicity many researchers fail to realize the role is becoming mainstream. 
It is the authors’ experience that upon learning of the RSE role, graduate students often become excited at the prospect of doing RSE work. 
If potential qualified individuals are not even aware that a role exists, they will not be searching for openings and therefore fail to enter the RSE pipeline.

Former researchers from backgrounds other than computer science (CS) are frequently faced with learning software development on their own, without formal training or mentorship. 
While this can be extremely effective, it can also leave gaps in knowledge unknown to the individual. 
It can also lead to imposter syndrome.

Many university RSE positions could be viewed as less desirable to other career opportunities, often seen by primary faculty as a lesser position that is not doing research, only supporting it and lacking the possibility of tenure. 
Positions are often grant funded, and therefore limited to a fixed term. 
It has been our experience that many former researchers are interested in job stability and specifically not having to tirelessly pursue additional funding.

It can be scary for researchers with a PhD to leave their own discipline and learn other disciplines. 
Exposing the advantages of an RSE position, including the opportunities to use existing knowledge in new areas and to learn and contribute to new disciplines can overcome this.

\subsection{Early Career Computer Science}
As mentioned earlier, over 75\% of RSEs have a background in a non-CS field. 
That means, however, that nearly 25\% of RSEs have a background in CS, larger than any other single field. 
Here we discuss early career professionals with a formal CS education. 
We define early career as up to three years of post-undergrad working experience. 
Beyond three years of experience, we would consider an RSE with an undergrad degree as a different entity, for example, an industry software engineer (see next section). 

Considering a typical RSE requires a mix of research and software engineering, a computer science undergraduate degree is likely the best formal education to prepare for the software engineering perspective. Additionally, many graduating students aren’t yet sure of their long term career goals and seek to further their skills in some way, perhaps getting a masters degree, or as we have seen, working in academia as an RSE for a few years.

As with the previous entry point, undergraduate students are unlikely to be aware of the RSE career path and perhaps even with research in general. 
Without exposure to graduate level research, the concept of research software is more abstract.
Therefore they are unaware of the potentially interesting, societally-relevant, and intellectual challenging projects associated with research software engineering.

Like all entry-level and early career professionals, a new RSE needs a significant amount of supervision and guidance. 
The aspects unique to RSEs in this category (as opposed to all entry-level positions) that seem to cause the most problems stem from elements inherent to the research workflow. 
This includes (a) the often nebulous and vague requirements of research software, (b) the incentives and priorities of researchers, and (c) the variation and inconsistency between projects.

Some CS students may not learn software engineering, or how to apply it. 
In this case, computer science can be seen as similar to any other discipline. 
However, the number of computing courses likely means that these RSEs have had some exposure to some software engineering practices, particularly at universities that emphasize using common tools and practices to prepare their students for jobs in industry.

\subsection{Industry Software Engineers}
A growing entry point into the Research Software Engineering field is from those with experience working as a software engineer in a non-research environment, which we’ll call “industry.” 
Clearly there are research software engineers at companies and private industry working as part of a research project, however for this classification we’re going to use the term to refer to those who are developing software for applications other than research. 

Exposure to best practices, rigorous software engineering, and working as part of larger teams are some of the clear benefits of having industry experience. 
Much of this expertise is transferable to the RSE role. 
In particular, software consultants from industry who also have hands-on experience are particularly valuable, as they have done work similar to that of senior RSEs, though perhaps not in a research context.

The biggest challenge with RSEs moving from industry to a research environment is the stark contrast in culture, priorities, and development environment. 
Many, if not most, non-research organizations are driven by business profit and as a result, software development practices reflect the business need. 
Strict, often unwavering, best practices must be followed at all times. 
Software engineers often work on software teams, with ownership only over small pieces. 
Software engineers are often removed from clients and customers relying on project managers to relay requirements. 
In an academic setting where requirements are often vague and change quickly combined with timelines that have lulls in oversight then sudden fast paced needs around things like conference and journal deadlines, or annual reviews, can be frustrating to software engineers from industry. 
Another challenge is that of motivation. 
Many who take the route of getting a degree in CS are motivated not only by the interest of the field but by the high monetary salaries of industry positions. 
This is typically not the case in academia, however. 
On the flip side, academia provides a good work-life balance as compared to many industry positions especially those in startups or fintech, often very good benefits with significant vacation time, interesting cutting-edge problems that can potentially benefit all of humanity, and the ability to have direct input, leadership, and ownership of the work. 
A good number of RSEs who come from industry gladly make the trade, wanting to be more than a cog in a massive machine, but pursuing recognition and impact, making a difference with their work.

The authors’ have observed a tendency among some RSEs coming from industry to not fully appreciate key aspects of research, in particular, they often don't have or don't see the need for the equivalent of a literature review (i.e., knowing the current landscape, what has been done already, etc.). 
This awareness of the field, however, is critical in order to truly drive innovation and impact and not reinvent the wheel.

\subsection{Other Notable Entry Points}
We believe the three categories capture a significant fraction of the current entry points into the RSE profession. 
Others not specifically addressed include:
\begin{itemize}
    \item Recent M.S. graduate. 
    This is something of a mix between the first two main entry points: domain science and entry-level CS. 
    Individuals in this category have more experience and research expertise than those who are entry-level, but less science and research experience than a PhD graduate.
    \item Professional research staff (including domain scientist or computer scientist). 
    Individuals in this category are even more experienced and mature than new PhD domain scientists but are often so entrenched in the research process that they struggle with transition to prioritizing software engineering aspects.
    \item Data scientist/engineer - Data scientists often employ many research software engineering approaches. 
    Moving to a data-heavy RSE project can be a smooth and logical transition.
    \item Research computing facilitator or research systems administrator. 
    Those in these positions are often adjacent to both RSEs and researchers, giving them a clear understanding of research incentives and processes. 
    In some cases, individuals in both of these jobs may already have an element of research software engineering as part of their regular job duties. 
\end{itemize}

\section{Proposed Activities}
As the RSE profession expands and garners national and international attention, the need to build a diverse and prepared pipeline and RSE workforce is becoming increasingly important to the long-term health of the profession. 
With the identification of the main entry points, we can begin to address the needs of each demographic. 
Here we propose specific activities to facilitate the preparation and success of new RSEs.

\subsection{Awareness}
\begin{figure*}
\centerline{\includegraphics[width=26pc]{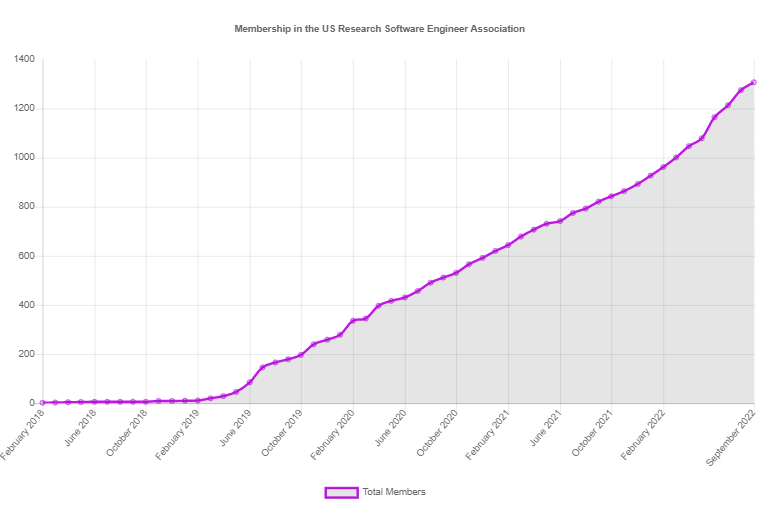}}
\caption{Number of members in the US-RSE organization (from \href{https://us-rse.org/join/}{https://us-rse.org/join/})
\label{us-rse-membership}}
\end{figure*}

While a relatively large number of individuals now identify as RSEs (see Figure~\ref{us-rse-membership}), we believe this is only the tip of the iceberg, with many software engineers in academia still unaware of this effort, without career paths, isolated and scattered across university projects, and living from grant to grant. 
This seems to stem from not only a lack of awareness from academic software engineers, but also from university leadership, despite the fact that those in leadership positions at major research universities are often working to address an ever increasing need for software, data, and computation as part of their research missions going forward. 
While most today acknowledge the need for better, bigger, more extensible and sustainable research software, the means to do this is still too often hiring students, encouraging researchers to do software development, tasking the campus IT staff with such development, or attempting to hire an external software firm and providing them with the software’s requirements if possible. 
As groups such as \href{https://us-rse.org}{the US Research Software Engineer Association (US-RSE)}\footnote{https://us-rse.org}, \href{https://researchsoft.org}{the Research Software Alliance (ReSA)}\footnote{https://researchsoft.org}, \href{https://academicdatascience.org/}{the Academic Data Science Alliance (ADSA)}\footnote{https://academicdatascience.org/}, etc. gain momentum, effort must be made to not only increase awareness of this RSE profession and how it benefits the research enterprise among software engineers in academia, but also among university leadership.

As outlined in the previous section, entry points into the RSE profession are varied, but because the RSE career path is still in its infancy, and there are no formal RSE education programs, word of mouth, professional networks, and in some cases luck often are the catalyst for future RSEs to learn of the profession. 
As a result, the potential pool for RSEs is often limited to those fortunate enough to stumble on the concept or those with significant professional networks. This limits the diversity and breadth of an RSE pipeline. 
We believe a concerted outreach effort by the existing community, with assistance and acceptance from the established research/educational community, is the best approach to expose potential future RSEs to the concept.

Given our past experience, outreach activities to students of all levels that both introduce the concept of RSEs and legitimize the career have been successful.
For example seminars and presentations, both in person and online, that explain RSE work with examples of actual RSE projects are frequently met with excitement and enthusiasm from students who were previously unaware.
The authors have received positive feedback from high school, undergraduate, and graduate students who have attended outreach talks.
Established RSE group leaders are in a unique position to give such presentations and should be encouraged and incentivized to do such outreach activities wherever possible. 

The US Research Software Engineer Association (US-RSE) was founded in early 2018, and in the following four years has grown to over 1300 members (see Figure~\ref{us-rse-membership}.) 
We believe that US-RSE is poised to make a significant impact on the future of the RSE profession. 
Many of the items we list here are directly tied to US-RSE’s mission to create a community, advocate for RSEs, provide resources in support of RSEs, and promote, encourage, and improve diversity, equity, and inclusion within the RSE profession.

US-RSE is part of the larger global RSE community, which started with the UK Research Software Engineers Association, which has now turned into the Society of Research Software Engineering. 
US-RSE, the Society, and other national and multinational RSE organizations work together on common activities, including on RSE recognition and promotion. 
In addition, the Research Software Alliance and national organizations that are working towards software sustainability, such as the \href{https://www.software.ac.uk/}{UK Software Sustainability Institute (SSI)}\footnote{https://www.software.ac.uk/} and the \href{https://urssi.us/}{US Research Software Sustainability Institute (URSSI)}\footnote{https://urssi.us/}, work closely with RSE organizations, as they understand that having a strong RSE community benefits the overall research software community.

National RSE associations, such as US-RSE, are increasingly positioned to organize and publicize outreach activities that bring awareness to students and other early career individuals.
US-RSE should continue to produce events such as early career panels, publish examples of RSE work, and facilitate sharing of outreach activities and approaches.  

\subsection{Training \& Education}
There is growing recognition amongst the scientific community for what are being called cyberprofessionals, of which RSEs are under. 
A major step on this front occurred in 2021 with the inclusion of a Cyberprofessional mentorship plan as part of proposals such as the CSSIs, requesting information as to how such staff are mentored, maintained, etc. 
Additionally, solicitations are more and more emerging around the training of cyberprofessionals, such as the CyberTraining call, with efforts spanning REU opportunities for students to work with experienced cyberprofessionals as an introduction to this area, to supporting the training of researchers and software engineers in particular areas with regards to emerging technologies.  The CyberAmbassadors effort for example, funded through the CyberTraining program for nearly 5 years now, has been very useful as a tool for introducing new cyberprofessionals to successfully working within an academic setting.  However, while we are beginning to see training and education opportunities for RSEs, there is still much more to be done as described next.

\subsubsection{(Research) Software Engineering}
To date, there are no formal accredited courses or programs in research software engineering. 
As mentioned above, new RSEs coming from a domain science often lack exposure to rigorous training or education on software engineering concepts. 
Easy access to targeted training material, workshops, and courses would enable those coming from non-CS backgrounds to fill the gaps in their knowledge and accelerate their self learning. 
While a number of software engineering bootcamps and online courses exist, few target research software specific issues and concepts at a level necessary for adequately preparing RSEs. 
\href{https://bssw.io/}{Better Scientific Software}\footnote{https://bssw.io/}, \href{https://intersect-training.org/}{INTERSECT}\footnote{https://intersect-training.org/}, and \href{https://coderefinery.org/}{CodeRefinery}\footnote{https://coderefinery.org/} are examples of projects and initiatives making progress in this area, but more work is needed. 

Because technical preparation is a major challenge for multiple RSE entry points, a formal degree program, for example, an MS in research software engineering would not only help domain scientists and recent computer science graduates gain much needed experience, but would also provide visibility and awareness of the RSE career path. 
It is our opinion that an MS in research software engineering would be a popular degree program and teach skills applicable for professions other than just research software engineering. 
In addition to RSEs, academic researchers who write software and industry software engineers would be reasonable and appropriate careers for graduates from a MS program. 

\subsubsection{Research fundamentals and domain science}
To address the need for learning research fundamentals for those without a domain science research background, targeted domain-specific programs would teach those new to the field the basics of research incentives, culture, and processes. 
This would not be meant to recreate a graduate degree, but rather prepare new RSEs to acquire the skills necessary to self-learn the elements of the domain in order to collaborate effectively with domain researchers.
More work is needed in this area both to identify relevant topics as well as to develop and deliver such domain-specific educational programs.
This is another area where national RSE associations, such as US-RSE, can and should take a leadership role in providing a forum for focused discussion and the creation of an educational framework that multiple domains could leverage.

\subsubsection{Software development vs. maintenance \& reuse} Because aspects of academia encourage creativity and innovation, there is a tendency at times to think any self-devised solution is novel by default and will solve a breadth of scientific challenges without first looking at what others have done (a time consuming task in and of itself).

Specifically with research software, a common problem is the endless reinventing of the wheel. 
This leads to terms such as ``yet another workflow system'' or ``yet another data management portal,'' systems that have most of the same capabilities as those that already exist, but one was made by biologists while another was made by geoscientists who were unaware of the others' work. 
Research software engineers need to have a mindset of starting with a literature/software review, not only on the science side, but also on the research software/cyberinfrastructure side, to understand if there is really a need to develop new software, or if it would be better to use existing software, perhaps adding a small number of features to it.

\subsubsection{Internship \& mentorship programs}
Our experience with our existing RSEs has been that the fastest and most effective way for new and aspiring RSEs to learn the necessary skills and knowledge is through formal and informal internship and mentorship programs. 
Even if more formal training avenues exist, Research Software Engineering is still a craft that needs practice and exposure. 
By working with a more experienced RSE mentor, a new or aspiring RSE can quickly learn applicable best practices and techniques, both in terms of technical skills as well as those social skills useful in contributing to team projects. 
This real world experience is effective and frequently transferable to other projects and/or domains.

Organizations with large RSE groups can support such programs internally, however, small programs with only a handful of RSEs, or even just one RSE, will struggle to provide internship and mentorship opportunities. 
It is our hope that the professional community, perhaps through national organizations such as US-RSE, can provide an accessible avenue for students and early career professionals to match with an internship or mentorship opportunity, even if it’s located at a different institution or organization.

\subsection{Professional support}
In the 10 years since the term RSE was coined, the profession has grown considerably and despite the work that remains, awareness has increased. 
The profession, however, still struggles with stability and long-term career prospects resulting from its relative nascency in the research ecosystem.
If our goal is to attract and retain talented and skilled professionals, we must make the profession desirable and appealing to join and remain in, not just for early career professionals, but for mid- and late-career professionals as well. 
This means having a clear and transparent career path with advancement opportunities. RSE position terms should be permanent and open-ended, providing RSEs with desirable and stable employment. 
Because it’s human nature to want to be valued and recognized for one’s work, the entire research community can help by giving credit for RSEs' contributions to research. 
Additional credit and recognition approaches, such as providing mechanisms for crediting software contributions, making software citable, and regularly citing software used \cite{smith}, will help raise the stature of RSEs and recognize the increasingly important contributions they make.  
This, in turn, will help attract and retain our best RSEs. 

\section{Conclusion}
Over the past 10 years, the idea of Research Software Engineers (RSEs) has been created and developed to recognize and promote a type of work that has long existed in scholarly research but has not had a uniform name or description. 
The RSE movement has built community among RSEs, many of whom didn't know they were RSEs initially, and has also built community awareness, leading to RSE associations in multiple countries, and RSE groups in many universities.

These formal groups have equally formal management structures, and one of the largest challenges for those who lead these groups is how to staff them. 
This involves finding qualified staff, either with the right software engineering skills or with other relevant knowledge and the ability to learn these skills. 
Additionally, to build this workforce, it's important for the possibility of an RSE career to be widespread, so that potential future candidates both are interested and develop a set of the necessary skills.

This then leads to a definition of how RSEs enter the field today, and what training is needed. 
RSEs generally come from one (or more) of three types of background: CS graduates, typically with an undergraduate or masters degree; PhD graduates, typically from a computational or data intensive field; and software engineers with experience in industry. 
For each, we have discussed the advantages and disadvantages they bring to an RSE position, as well as how we might better expose them to the possibility of such a position, and how we might educate and train them for it.

We have suggested a number of community needs, such as more activities to increase awareness, train and educate future RSEs, and to support current RSEs through increased recognition. 
We believe that national organizations (for example, US-RSE in the United States) can and should play a significant role in both sponsoring such activities and providing the documentation and resources necessary to support the organization of local activities.

\subsection{Acknowledgment}

We would like to thank the overwhelmingly motivated and positive members of the both the US and international RSE community for openly collaborating and supporting any and all RSE efforts. 
A special thanks to the early members of the The Society for Research Software Engineering (formerly the UK RSE Association) for starting ten years ago what would become an international RSE movement.
We also want to thanks our own organizations, Princeton University and NCSA/University of Illinois, for their support of the RSE movement and our RSE groups.

\begin{IEEEbiography}{Ian A. Cosden}{\,}is Director for Research Software Engineering for Computational and Data Science at Princeton University, in Princeton, NJ.
He earned a Bachelors in Mechanical Engineering from University of Delaware, an M.S. in Mechanical Engineering from Syracuse University, and a Ph.D. in Mechanical Engineering from the University of Pennsylvania.
At Princeton, he leads a team of Research Software Engineers (RSEs) who complement multiple traditional academic research groups by offering embedded, long-term software development expertise. 
Additionally, he is the current and founding chair of the Steering Committee for the US Research Software Engineer Association (US-RSE). 
Contact him at icosden@princeton.edu.
\end{IEEEbiography}

\begin{IEEEbiography}{Kenton McHenry} is Associate Director for Software at the National Center for Supercomputing Applications (NCSA) at the University of Illinois at Urbana-Champaign.  He received his B.S. in Computer Science from California State University San Bernardino, and Ph.D. in Computer Science/Artifical Intelligence at the University of Illinois at Urbana-Champaign.  At NCSA he leads a team of RSEs supporting software development, data sharing, and analytics needs across the Illinois campus as well as externally.  Contact him at mchenry@illinois.edu.
\end{IEEEbiography}

\begin{IEEEbiography}{Daniel S. Katz} is Chief Scientist at the National Center for Supercomputing Applications (NCSA) and Research Associate Professor in Computer Science, Electrical and Computer Engineering, and the School of Information Sciences (iSchool), at the University of Illinois at Urbana-Champaign. He received his B.S., M.S., and Ph.D degrees in Electrical Engineering from Northwestern University, Evanston, Illinois, in 1988, 1990, and 1994, respectively. Dan studies policy issues, including citation and credit mechanisms and practices associated with software and data, organization and community practices for collaboration, and career paths for computing researchers. He is a senior member of the IEEE, the IEEE Computer Society, and ACM, co-founder and current Associate Editor-in-Chief of the Journal of Open Source Software, co-founder of the US Research Software Engineer Association (US-RSE), and co-founder and steering committee chair of the Research Software Alliance (ReSA). Contact him at d.katz@ieee.org.
\end{IEEEbiography}


\begin{thebibliography}{1}

\bibitem{baxter}
R. Baxter,  N. C. Hong, D. Gorissen, J Hetherington, I. Todorov. ``The Research Software Engineer.'' In Proceedings of the Software Sustainability Institute Collaborations Workshop, 2012.



\bibitem{katz-1}
D. S. Katz, K. McHenry, C. Reinking, and R. Haines, ``Research Software Development \& Management in Universities: Case Studies from Manchester's RSDS Group, Illinois' NCSA, and Notre Dame's CRC,'' IEEE/ACM 14th International Workshop on Software Engineering for Science (SE4Science). IEEE, May 2019. doi: 10.1109/SE4Science.2019.00009.

\bibitem{katz-2}
D. S. Katz, K. McHenry and J. S. Lee, ``Research Software Sustainability: Lessons Learned at NCSA,'' Proceedings of the 54th Hawaii International Conference on System Sciences. University of Hawaii, Jan. 2021. \url{http://hdl.handle.net/10125/71494}


\bibitem{princeton-rse-expansion}
Princeton University, ``Princeton bets big on research software engineering.'' [Online]. Available: \href{https://researchcomputing.princeton.edu/news/2022/princeton-bets-big-research-software-engineering}{https://researchcomputing.princeton.edu/news/2022/\break
princeton-bets-big-research-software-engineering}. 

\bibitem{hettrick}
S. Hettrick et al., International RSE Survey 2022. Zenodo, 2022. doi: 10.5281/ZENODO.6884882.

\bibitem{smith}
A. M. Smith, D. S. Katz, and K. E. Niemeyer, “Software citation principles,” PeerJ Computer Science, vol. 2. PeerJ, p. e86, Sep. 19, 2016. doi: 10.7717/peerj-cs.86.

\end{thebibliography}
\end{document}